# Direct observation of ordered configurations of hydrogen adatoms on graphene

*Chenfang Lin[1†], Yexin Feng[2,3†], Yingdong Xiao[1†], Michael Duerr[4], Xiangqian Huang[1], Xiaozhi Xu[1], Ruguang Zhao[1], Enge Wang[2,3]\*, Xin-Zheng Li[1,2]\* and Zonghai Hu[1,2]\**

[1]State Key Lab for Mesoscopic Physics, School of Physics, Peking University, Beijing 100871, China

[2]Collaborative Innovation Center for Quantum Matter, Beijing, China

[3]International Center for Quantum Materials, School of Physics, Peking University, Beijing, China

[4]Institute of Applied Physics, Justus Liebig University Giessen, 35392 Giessen, Germany

[†] These authors contributed equally to this work.

\* Corresponding authors: egwang@pku.edu.cn, xzli@pku.edu.cn, zhhu@pku.edu.cn

**ABSTRACT** Ordered configurations of hydrogen adatoms on graphene have long been proposed, calculated and searched for. Here we report direct observation of several ordered configurations of H adatoms on graphene by scanning tunneling microscopy. On the top side of the graphene plane, H atoms in the configurations appear to stick to carbon atoms in the same sublattice. A gap larger than 0.6 eV in the local density of states of the configurations was revealed by scanning tunneling spectroscopy measurements. These findings can be well explained by density functional theory calculations based on double sided H configurations. In addition, factors that may influence H ordering are discussed.

# Introduction

Adsorbates can vastly change the atomic and electronic structures of two-dimensional (2D) materials. Therefore they are major candidates for property tuning and producing new materials especially if they form ordered configurations.[1-4] Hydrogen/graphene has become a model system because graphene is the most studied 2D material and hydrogen is the simplest adsorbate species. This system also finds practical relevance in hydrogen storage,[5] $H_2$ formation in the universe,[6] Tokamak wall erosion,[7] and recently, graphene based electronic and spintronic devices.[8-11] Concerning this most recent development, ordered configurations have received great attention because they are closely tied to tuning of graphene properties including large band gap opening and formation of specific magnetic orders, both of which are highly desirable in potential applications. Hydrogenation can induce magnetic moment, enhance the spin-orbit coupling and change the magnetic properties of graphene.[12-15] An imbalance in the H occupancy of the two sublattices can generate ferromagnetism in graphene. Ordered H distributions rather than random ones are better for realizing this imbalance. Pristine graphene is a zero gap semiconductor with exotic properties.[16,17] However, a band gap is needed in devices such as field effect transistors. Hydrogenation has emerged as a promising method to generate an energy gap in graphene.[1,2,8-10,18,19] Elias *et al.* demonstrated a disorder-induced metal-insulator transition and a mobility gap after hydrogenation of graphene on $SiO_2$ substrates.[2,18] After double-sided hydrogenation of suspended graphene, they observed structural deformation in the carbon lattice, a sign of graphane-like structures.[1,2] Balog *et al.* studied the case of patterned hydrogen adsorption on graphene moiré superstructures on Ir(111). They proposed a graphane-like ordered structure in the hydrogenated area involving C-Ir bonding to explain the coverage-dependant energy gap as a confinement effect.[8,9] To achieve practical on/off performance in devices at room temperature, a true band gap is highly beneficial. This requires that the H adatoms in ordered configurations.

Many ordered structures of hydrogenated graphene have been proposed, including double sided and single sided ones, with the calculated band gap width depending on the respective H coverage.[1,8,20-22] For example, theoretical work by Sofo *et al.* showed that graphane, the fully hydrogenated graphene monolayer with the H atoms bonding to carbon on both sides of the plane in an alternating manner, possesses a band gap larger than 3.5 eV.[1] However, although these ordered structures have received broad attention, none has been observed directly and the possibility of ordered hydrogenation of graphene remains in doubt to date. To the contrary, pairing and clustering of H adatoms have been observed in single sided hydrogenation, and the H atoms appear to be disordered.[23-27]

To lift the above uncertainty, here we report direct imaging of several ordered configurations of H adatoms on graphene by scanning tunneling microscopy (STM). The H atoms in the configurations exhibit apparent sublattice selectivity and tiny deviations from the exact atop-of-carbon positions. Scanning tunneling spectroscopy (STS) measurements of the configurations showed a larger than 0.6 eV gap in the local density of states (LDOS). These findings can be well explained by our density functional theory (DFT) simulations based on models of double sided H configurations. Factors that may influence H ordering and large area formation are also discussed.

**Results and Discussion**

Graphene samples grown on copper foil by the standard chemical vapor deposition (CVD) method were dosed with a hydrogen beam generated in a hot capillary heated to >2900 K. The typical H dosage was at a flux of ~1 × $10^{11}$ cm$^{-2}$/s for 2 hours. STM was done in UHV at 77K. DFT calculations were performed using the VASP code with the Perdew-Burke-Ernzerhof (PBE) functionals.[28,29] The projector augmented wave (PAW) potentials were used with a 700eV plane wave cutoff energy. Binding energies of Configuration A, B and C were calculated employing a 6 × 6 graphene supercell with a 3 × 3 × 1 mesh of the Monkhorst-Pack *k* points. The H diffusion barriers were calculated using the climbing image nudged elastic band (NEB) method[30].

Figure 1 shows the Raman spectra of the samples. The intensity of the defect-associated graphene D band ($I_D$) increases with the H dosage, indicating H uptake. Meanwhile the G band broadens and the D' band emerges. From the $I_D/I_G$ ratio, the H coverage of the sample corresponding to the blue spectrum is estimated to be 4%.[31] We note that the H coverage is not uniform over the surface. Some areas appear clean of H while others with high H coverage.

Atomically resolved STM images of some ordered H configurations and their corresponding schematic diagrams are shown in Figure 2. H adatoms appear as the bright spots in STM. Configuration A in Fig. 2a exhibits a $\sqrt{3}\times\sqrt{3}$ / R30° structure with respect to the graphene lattice (for convenience we call this configuration "graphine"). A closer look at Fig. 2a reveals that the H atoms appear a little bit off the "on-top" positions of the occupied carbon sites, upwards along the C-C bonds. Configuration B in Fig. 2c can be viewed as two staggered sets of "graphine", displaced by a lattice constant of graphene. However, there is a subtle deformation of the hexagonal shape as illustrated in Fig. 2d, |AB| × 2 > |CD|. Configuration C in Fig. 2e can be considered as three interleaving sets of "graphine", displaced to each other by the lattice constant (every other carbon site appearing bright). In our experiments, Configuration B and C were more often observed than Configuration A. Extended area larger than a few nm$^2$ of one single configuration was hard to obtain. Instead, mixture of the three configurations was often seen, as shown in Fig. 2h. Interestingly, almost all the bright spots appear to occupy the same sublattice. Though it seems that some of the graphene honeycomb cells are deformed and the H ordering is not perfect, significant ordering is still found in the corresponding fast Fourier transformation (FFT) shown in Fig. 2i. The diffraction spots marked by the green circles correspond to the graphene lattice and Configuration C. Spots in the blue circles correspond to Configuration A and B (containing the $\sqrt{3}\times\sqrt{3}$ / R30° structure).

The STM tip could disturb the configurations from time to time, especially with large values of bias and tunneling current. Fig. 3a and 3b are STM scans of the same area with a time interval of 10 minutes. The white circles highlight one of the disappearing bright spots in the later scan. Notably, this verifies that the bright spots seen here are indeed H adatoms, not the scattering patterns caused by nearby defects.[32]

To investigate the effects of these ordered H configurations on the band structure of graphene, STS measurements were taken and significant changes in the LDOS were observed. Fig. 3d and 3e are typical STS results of ordered hydrogenated regions. There is a gap in the LDOS ranging from 0.6 eV to 1.2 eV. As a comparison, STS results of clean graphene regions (Fig. 3c) show the well known Dirac point feature. There are hydrogenated regions with a gap larger than 2 eV in the LDOS, however it is hard to achieve high spatial resolution in these regions to identify any ordering. This is one of the reasons why the ordered configurations have been elusive for so long. Another possible reason is the easy displacement of the H adatoms by STM tips.

Extensive studies have shown that the single sided *meta*-dimers (H occupying two next-nearest-neighbor sites in one hexagon) are energetically much less stable than the *ortho*-dimers (two nearest-neighbor sites) and *para*-dimers (two sites on the opposite corners of one hexagon),[23] so is the *chair*-graphone configuration (H atoms fully occupying one sublattice).[33] Therefore by first sight our observation of the ordered H atoms (bright spots) occupying the same sublattice is puzzeling. However, since graphene is only weakly bounded to the Cu substrate and intercalation can happen,[34-37] another possible scenario is that the bright spots correspond to double sided *ortho*-dimers (one H above and one below the graphene plane on two nearest neighbor sites) and only the top sided H atoms are seen in STM. To check the validity of this assumption, we carried out DFT calculations. Good agreements with our experimental results were found. The binding energies per H atom for several structures on freestanding monolayer graphene are summarized in Table 1. Indeed, our calculations show that the double sided

*ortho*-dimer is the most stable dimer configuration, in agreement with recent reports.[38] For Configuration A, we first considered single-side hydrogenation (denoted as s-A in Table 1) with H coverage of 1/6. The resultant H binding energy is even lower than that of isolated H monomers. This indicates that although the single sided model will give the look of Configuration A, it is energetically unfavorable. For Configuration A with double sided *ortho*-dimers (denoted as d-A in Table 1), the resultant H binding energy is much higher than all isolated dimer configurations. The binding energies of double sided Configuration B and C are even higher. This indicates that all the three double sided configurations are energetically more favorable than isolated H monomers or dimers.

Figure 4 shows the 2D partial charge density distributions of double sided Configuration A, B and C together with the atomic structures. A close look at Fig. 4a reveals that because of the existence of the bottom sided H atoms at the nearest neighbor carbon sites, the top sided H atoms (the bright spots) appear to move toward the center of the C-C bonds, in contrast to the on-top positions of H monomers. Such a phenomenon is indeed observed in Fig. 2a as mentioned above. Moreover, Fig. 4b reveals that similar small deviations from the on-top positions lead to apparent distortion of the hexagonal shape of the bright spots, so that |AB| × 2 > |CD|, again in excellent agreement with the STM image Fig. 2c. These detailed agreements further support the double sided picture of the ordered H configurations.

Figure 5 shows the band structure and DOS for the three configurations. A gap exists in all of them. Configuration A possesses a gap of 0.6 eV, while both Configuration B and C have a wide band gap larger than 3.4 eV. Our experimentally measured gap (>0.6 eV) in the LDOS is already substantialand depends on the size of the ordered area and the mixture of the three configurations in the area. (We noted above that in regions with a gap larger than 2 eV, no atomic resolution was achieved). To obtain a relatively uniform gap throughout the sample, formation of large size ordered area is needed.

Because of the sublattice compatibility on each side of the graphene plane, all three configurations can be considered as partial formation *en route* to a complete *chair* graphane structure which has the highest binding energy per H atom among all configurations. Besides the binding energy, there are various dynamic processes including adsorption, diffusion, abstraction, and desorption that can affect H ordering.[3,7,39-41] Since the kinetic energy of the H atoms generated in our hot capillary (>2900 K) is higher than the reported ~0.2 eV adsorption barrier on clean graphene surface,[24,42] the sticking sites should be largely random. Therefore adequate H diffusion should be important for ordering. H atoms can find energetically favorable positions through diffusion. Because H is the lightest atom, quantum treatment of the proton movement has given a substantial diffusion rate of ~8 $s^{-1}$ for isolated H atoms at room temperature, 20 times of the value obtained by classical molecular dynamics method.[40] Indeed, small flux and long adsorption and diffusion time in our experiments resulted in better ordering than the opposite conditions. On the bottom side of graphene, the H atoms prefer to diffuse on the copper surface with a barrier of ~0.15 eV,[43] much faster than on graphene with a barrier of ~1 eV.[44] Then, the transfer of H atoms from the Cu surface to graphene becomes critical. Our climbing image nudged elastic band (NEB) simulations show that a topside-adsorbed H atom will greatly promote this process, lowering the transfer barrier from 1.21 eV (without the topside H) to 0.31 eV. This results in the preferential sticking configuration - the double sided *ortho*-dimer. The H-assisted sticking process can go on and lead to the "growth" of the ordered configurations. We note that because the permeation barrier is prohibitively high for H to go through an intact graphene layer,[45] the intercalation needs to start at defect sites such as grain boundaries.[37,46] For this reason, suspended graphene may better facilitate the double sided adsorption and similar H-assisted preferential sticking mechanisms should still work.

Desorption and abstraction will affect the H saturation coverage.[7] Intertwined with adsorption and diffusion, they can have profound effects on ordering. Other factors including defects, surface curvature and graphene-substrate interaction can further complicate the processes.[8,9] Systematic experimental and theoretical investigations of all these processes are much needed but beyond the scope of the current work. Nevertheless, our demonstration of the existence of ordered H configurations should inspire the design and synthesis of new ordered adsorbate structures to form new graphene based functional materials. Regarding magnetism, the three double sided configurations are nonmagnetic because the magnetic moment induced by H adsorption is suppressed in the *ortho*-dimer unit. However, if we can keep the top sided H atoms and remove the bottom sided ones, the configurations become ferromagnetic because of the sublattice selectivity. Recent studies suggest that by using intense ultra-short *p*-polarized laser pulses with an asymmetric time envelope, selective removal of H atoms from one side of graphene can be achieved.[47]

In summary, ordered configurations of H adatoms on graphene was directly observed by STM. The H atoms in the configurations exhibit apparent sublattice selectivity and tiny deviations from the exact atop-of-carbon positions. STS measurements of the configurations showed a larger than 0.6 eV gap in the LDOS. These findings can be well explained by our DFT simulations based on models of double sided H configurations. Our results are relevant to band gap opening and obtaining specific magnetic orders in graphene.

**Acknowledgments**

Z. H. thanks the NBRP of China (Grant 2012CB921300) and the Chinese Ministry of Education for financial supports. E. W., X. Z. L. and Z. H. thank the National Natural Science Foundation of China (Grant 11074005, 91021007 and 11275008) for financial supports.

**Figure captions**

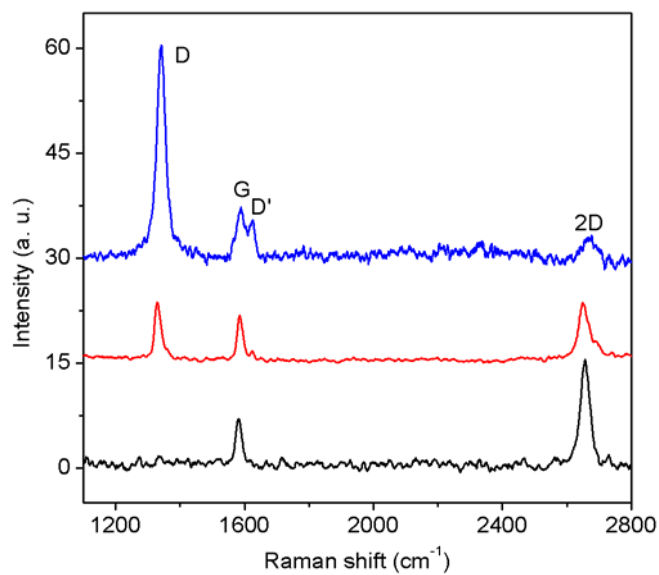

**FIGURE 1**. Raman spectra of as-grown graphene (black), graphene exposed to a lower H dosage (red) and a higher dosage (blue) with an excitation wavelength of 633 nm. The intensity is renormalized to the G band intensity. The background caused by the substrate was subtracted and the spectra were vertically displaced for clarity. The increase of $I_D$ and $I_D/I_G$ ratio with increasing H dosage indicates H uptake.

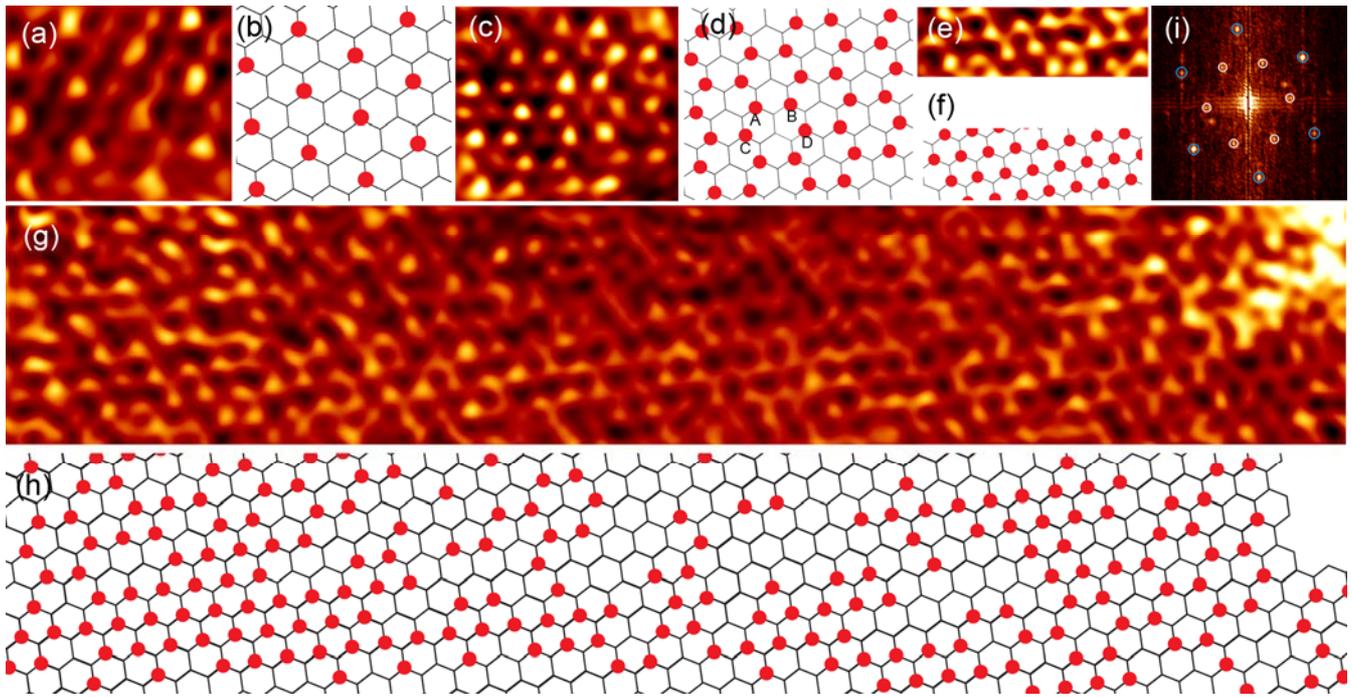

**FIGURE 2**. STM images showing ordered configurations of H adatoms. **a. b.** Configuration A (graphine). The H atoms appear a little bit off the "on-top" positions of the occupied carbon sites, moving upwards along the C-C bonds. (1.4 nm × 1.2 nm scan size). **c. d.** Configuration B. |AB| × 2 > |CD|. (1.8 nm × 1.6 nm); **e. f.** Configuration C. (1.9 nm × 0.6 nm). **g. h.** mixing of the three structures mentioned above. **i.** Fast Fourier transformation of g. The spots corresponding to the graphene lattice and the $\sqrt{3} \times \sqrt{3}$ / R30° hydrogen structure are marked by the blue and wihte circles, respectively. Scanning conditions: tunneling current 0.1 nA, sample bias -1 V.

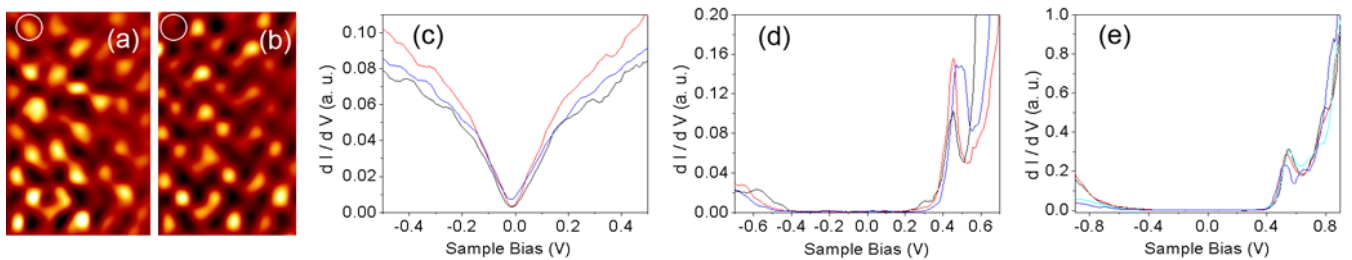

**FIGURE 3**. **a. b.** Consequent STM scans of the same area. The white circles highlight one of the disappearing H atoms. **c.** STS results taken at different points in clean regions (without H atoms), showing the Dirac point. **d. e.** STS in ordered regions, showing a gap ranging from 0.6 to 1.2 eV in the LDOS in different regions. Set point: I = 0.2 nA, sample bias V = -1 V, modulation voltage (peak-to-peak value) ΔV = 10 mV.

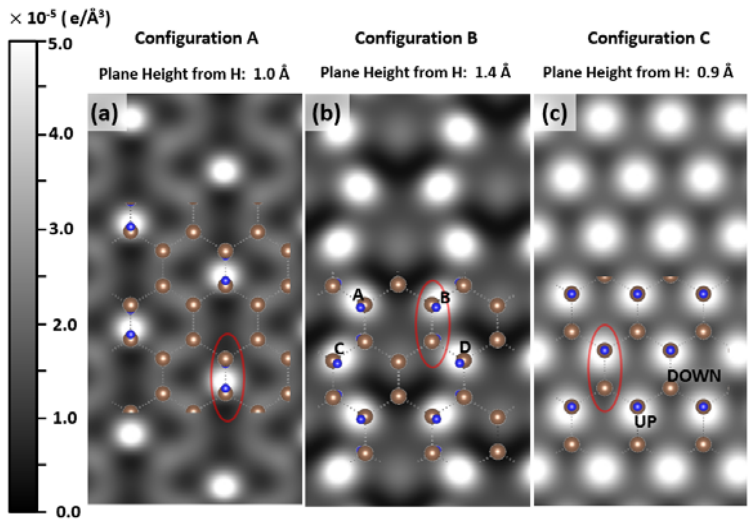

**FIGURE 4**. DFT simulation results of the 2D contours of the electron partial charge density for Configuration A, B and C. The brown (blue) balls mark the C (H) atoms. The double sided *ortho*-dimer units are indicated by the red ovals. In **a.**, the H atoms appear a little bit off the "on-top" positions of the occupied carbon sites, moving upwards along the C-C bonds. In **b.**, $|AB| \times 2 > |CD|$.

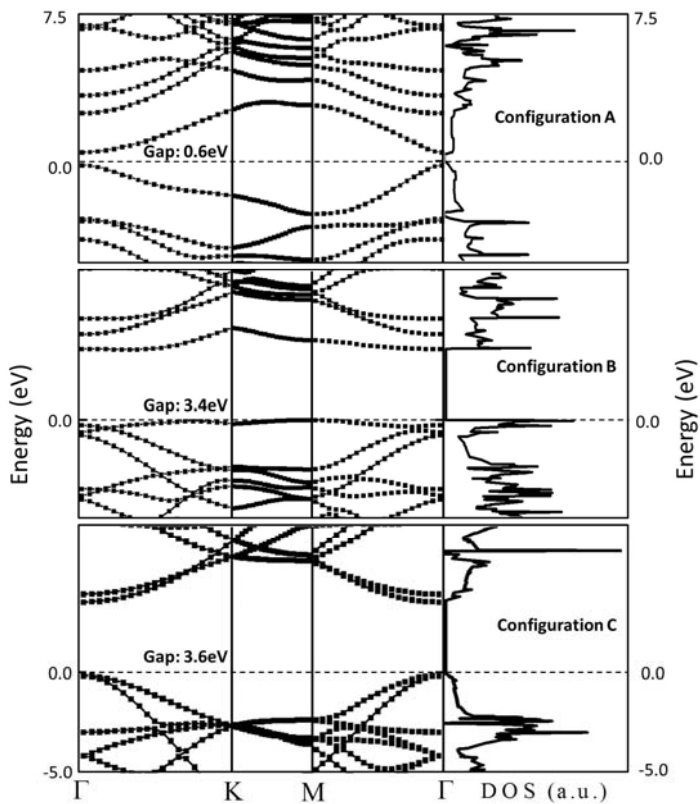

**FIGURE 5**. DFT results of the band structures and DOS for Configuration A, B and C.

**TABLE 1.** H binding energies per H atom for several structures. 's' denotes single sided and 'd' denotes double sided. "*ortho*", "*para*", and "*meta*" denote the different dimer configurations. A, B, C stand for the ordered structures observed in our STM experiments.

| Structure | monomer | d-*ortho* | d-*para* | d-*meta* | s-*ortho* | s-*para* | s-*meta* | s-A | d-A | d-B | d-C |
|---|---|---|---|---|---|---|---|---|---|---|---|
| $E_b$(eV) | 0.83 | 1.66 | 1.27 | 0.76 | 1.38 | 1.35 | 0.76 | 0.76 | 1.82 | 2.22 | 2.45 |